\documentclass[
aip,
amsmath,amssymb,reprint,floatfix
]{revtex4-1}

\usepackage{physics}
\usepackage{verbatim}
\usepackage{natbib}
\usepackage{placeins}
\usepackage{adjustbox}
\usepackage{tikz}
\usepackage{soul}
\usepackage{subcaption}
\usepackage{cleveref}
\usepackage{graphicx}
\usepackage{float}
\usepackage{xcolor}

\Crefname{equation}{Eq.}{Eqs.}
\Crefname{figure}{Fig.}{Figs.}
\DeclareMathAlphabet{\mathpzc}{OT1}{pzc}{m}{it}
\DeclareMathAlphabet\mathbfcal{OMS}{cmsy}{b}{n}
\begin{document}
\title{Nonlinear regimes of the electron cyclotron drift instability in Vlasov simulations}
\author{Arash Tavassoli}
	\email{art562@usask.ca}
	\author{Andrei Smolyakov}
	\author{Magdi Shoucri}
	\affiliation{Department of Physics and Engineering Physics, University of Saskatchewan,  Saskatoon,  Saskatchewan, Canada S7N 5E2}
	\author{Raymond J. Spiteri}
	\affiliation{Department of Computer Science, University of Saskatchewan,  Saskatoon, Saskatchewan, Canada S7N 5C9}
	\date{\today}

\begin{abstract}
    We report on a novel investigation of the nonlinear regime of the electron cyclotron drift instability using a grid-based Vlasov simulation. It is shown that the instability occurs as a series of cyclotron resonances with the electron beam mode due to the $E\times B$ drift. In the nonlinear regime, we  observe condensation of fluctuations energy toward the lowest resonance mode and below, i.e., an inverse energy cascade. It is shown that the characteristics of the nonlinear saturation state remain far from the ion-sound regime.
\end{abstract}

\maketitle

 It is widely accepted that the anomalous transport across the magnetic field in partially magnetized plasmas results from turbulent electrostatic fluctuations. The exact mechanism of such  fluctuations in various conditions, in particular those typical for Hall thrusters, is still a subject of debate. In recent years, the electron cyclotron drift instability (ECDI), or simply the electron drift instability (EDI), has attracted a great deal of attention as a mechanism of the anomalous transport in partially magnetized plasmas, especially in the strong electric field regions, where other effects, e.g.,~density and magnetic field gradients, are less important \cite{janhunen2018nonlinear,janhunen2018evolution,lafleur2016theory1,croes20172d}. This instability is also of interest for space physics, in particular, as a dissipation mechanism for collisionless shock waves \cite{muschietti2013microturbulence}.    

The ECDI is a reactive instability driven by the relative drift velocity between electrons and ions in partially magnetized plasmas with crossed E and B fields. An electric field $\vb{E}_0$ applied across the magnetic field generates an $E\times B$ drift velocity $\vb{v}_d=\frac{\vb{E}_0\times \vb{B}_0}{B^2}$ of the bulk  electrons (the ions are assumed to be unmagnetized). The instability occurs when the resonances of Bernstein-type (cyclotron) modes and the ion-sound mode become possible due to the Doppler frequency shift due to the electron $E\times B$ drift. Under the conditions in Hall thrusters, where the magnetic field $\vb{B}_0$ and the electric field $\vb{E}_0$ are applied in the radial and axial directions, respectively, the fluctuations propagate in the azimuthal direction \cite{boeuf2017tutorial}. Experiments \cite{tsikata2009dispersion,brown2019spatial} report observations of such small-scale azimuthal fluctuations in the acceleration region of Hall thrusters are presumably responsible for anomalous axial electron transport.

The linear regime of the ECDI is well understood based on the linear dispersion relation \cite{gary1993theory,forslund1970electron}. However, understanding of the nonlinear regimes remains elusive.
In part, the understanding is obscured by the results of the linear theory that for  finite and sufficiently large values of the wave vector along the magnetic field,  $k_z v_{te} >\omega _{ce}$ ($v_{te}$ is the electron thermal velocity and $\omega _{ce}$ is the electron cyclotron frequency), the cyclotron resonances are smeared out by the electron motion along the magnetic field, and the instability is reduced to the ion-sound instability driven by the electron $E \times B$ beam. 

 It has been suggested that even for purely perpendicular propagation, when $k_z=0$ and the above linear effect is absent, the nonlinear resonance broadening due to the nonlinear diffusion may result in the overlapping resonances effectively demagnetizing the electron response.
A nonlinear theory of the ECDI based on the resonance broadening in the strong turbulence regime \cite{dupree1966perturbation,dum1969saturation} was proposed in Refs.~\onlinecite{lampe1971nonlinear,lampe1972theory,lampe1972anomalous}.  As a result, an initial strong ECDI instability would saturate and proceed further as a slow ion-sound instability, similar to the ion-sound instability in plasmas without a magnetic field. Such behavior was also demonstrated in earlier Particle-in-Cell (PIC) simulations    
\cite{lampe1971nonlinear}. At the same time, it was argued that in similar PIC simulations, properties of the ECDI remain unlike those of unmagnetized ion-sound instability \cite{forslund1972anomalous}. Comparison of the properties of ECDI with unmagnetized ion-sound instability in the context of the collisionless shock waves in space also showed significant differences \cite{muschietti2013microturbulence}.   

A quasi-linear theory based on the assumption of unmagnetized ion-sound turbulence was used to explain the anomalous mobility caused by the ECDI \cite{lafleur2016theory2}. In this approach, the anomalous current is calculated from the $E\times B$ drift, in the self-consistent electric field, i.e.,~$J_{ze}=\frac{\expval{n_e\vb{\tilde{E}}\times\vb{B}_0}}{B^2_0}$, where $n_e$ and $\vb{\tilde{E}}$ are the electron density and the self-consistent electric field and are calculated as for unmagnetized ion-sound turbulence. These results were validated against some particle-in-cell (PIC) simulations\cite{lafleur2018anomalous,croes2018effect,charoy2020comparison}.

Many PIC simulations have been performed recently to investigate the nonlinear regimes of the ECDI in various conditions \cite{janhunen2018nonlinear,janhunen2018evolution,lafleur2016theory1,croes20172d,ducrocq2006high,adam2004study,villafana20212d,charoy20192d,TaccognaPSST2019,asadi2019numerical}.  In Ref.~\onlinecite{janhunen2018nonlinear}, one-dimensional (1D) PIC simulations of the ECDI  are presented in the parameter regime close to the typical conditions of Hall thruster operation. Various aspects of the nonlinear behavior, such as the significant flattening of the distribution function from Maxwellian and the inverse cascade of electrostatic energy to low-$k$ modes, were revealed. It was shown that the criteria of Refs.~\onlinecite{lampe1971nonlinear,lampe1972theory} for electron demagnetization are not generally satisfied. Two-dimensional (2D) effects are also studied in a subsequent paper\cite{janhunen2018evolution}, where it is found that the nonlinear regime is affected  by the long-wavelength modified-two-stream modes and the low harmonics of the ECDI modes, so that the magnetic field remains important, unlike  the unmagnetized ion-sound instability. 

A weakness of the PIC method is the significant amount of noise it introduces in the simulations due to the finite number of macro particles. It is known that the noise of PIC simulations can significantly affect the overall physics of a problem, potentially changing the outcome \cite{tavassoli2021role}. An example of such effects was observed in PIC simulations of the heat transport in electron temperature gradient turbulence, which can be dominated by noise effects \cite{nevins2005discrete}. The noise of PIC simulations can be reduced by increasing the number of the macro particles, but this reduction comes at a large computational cost because the noise only decreases as the inverse square root of number of macro particles. In studies of the ECDI, there are concerns that the noise of PIC simulations can facilitate resonance broadening and transition to the ion-sound regime by enhancing the nonlinear diffusion effects \cite{lafleur2016theory1,janhunen2018nonlinear,kaganovich2020physics}. 

An alternative to the PIC method is based on the Vlasov equation, which is known to be free of the statistical noise introduced by the discrete nature of the macro particles in PIC simulations. In this Letter, we report on novel investigations of the ECDI using low-noise, grid-based Vlasov simulations in one spatial and two velocity dimensions. Although Vlasov simulations of plasma instabilities are well studied and widely used in various settings \cite{CalifanoJPP2016} including applications relevant to Hall thrusters \cite{HaraPSST2018}, to the best of our knowledge, this is the first attempt to specifically address the nonlinear regime of the ECDI using Vlasov simulations.  We show that the nonlinear stage of the mode growth exhibits several transitions between different regimes dominated by the growth of the low-$k$ modes, attaining increasing growth rates much larger than those of the high-$k$ modes.  An intense first cyclotron-resonance mode appears in the initial nonlinear stage, with even longer wavelength modes appearing in the later stages. Similar transitions are also observed in the spatial profile of the electron density. This behavior is a signature of the inverse cascade of the electrostatic energy towards low-$k$ modes. The wavelengths of these high-growth-rate, low-$k$ modes remain well below those of the ion-sound modes, which have a maximum growth rate of $1/\sqrt{2}\lambda_D$, where $\lambda_D$ is the Debye length of electrons. This discrepancy suggests that the nonlinear regime of ECDI cannot be explained by the ion-sound turbulence theory for an unmagnetized plasma. 

In our setup, we take $\vb{B}_0 = B_0 \hat{\vb{y}}$, $\vb{E}_0 = E_0 \hat{\vb{z}}$, and $\vb{\tilde{E}}=E_x\hat{\vb{x}}$. The code used in this study is a 1D2V Vlasov code that uses the well-known semi-Lagrangian method\cite{cheng1976integration,cheng1977integration,gagne1977splitting,magdi2009method}. A second-order operator splitting scheme is used in this method\cite{macnamara2016operator}. The Vlasov equation of the electrons and ions are split into three equations; the convection equation, the momentum balance equation in the  $\hat{\vb{x}}$ direction, and the momentum balance  equation in the $\hat{\vb{z}}$ direction, noting that the ions are unmagnetized and hence the last equation for ions is trivial.  Each equation is then solved using the method of characteristics with cubic-spline interpolation. The Poisson equation is solved, following the convection steps, using the fast Fourier transform. In our simulation, we use  parameters close to the typical operation regime of the Hall thrusters that were also used in the PIC simulations of Refs.~\onlinecite{janhunen2018nonlinear,janhunen2018evolution}; i.e.,~$E_0=200$ V/cm, $B_0=200$ G, ion mass $m_i=133.3$~u (Xenon), density $n_0=10^{17}\;\text{m}^{-3}$, electron temperature $T_{e0}=10$ eV, and ion temperature $T_{i0}=0.2$ eV. The length of the system is taken to be $L=4.456$ cm, and 2048 cells are used to resolve it,  giving a partial resolution of $0.29\;\lambda_D$. The velocity grids consist of $1200\times 1200$ cells with a resolution of $0.054\;v_{te}$ for the electrons and $200\times 200$ cells with a resolution of $10^{-4}\;v_{te}$ for the ions. Due to the low-noise feature of the Vlasov simulations, an initial perturbation is required to excite the instability. Accordingly, we perturb the initial densities as $n_i=n_e=n_0\qty[1+1.41\times 10^{-4}\cos (2\pi x/L)]$. The boundary conditions are periodic in physical space and open in velocity space. The time step in our simulations is $1.1\times 10^{-11}\; \text{s} \approx 0.2\;\omega_{pe}$. To confirm the validity of our results, we have also repeated the simulation with finer grids and smaller time steps, and the results were in good agreement.

\begin{figure}[htbp]
\centering
\captionsetup[subfigure]{labelformat=empty}
\subcaptionbox{\label{fig:growth_rates}}{\includegraphics[width=1\linewidth]{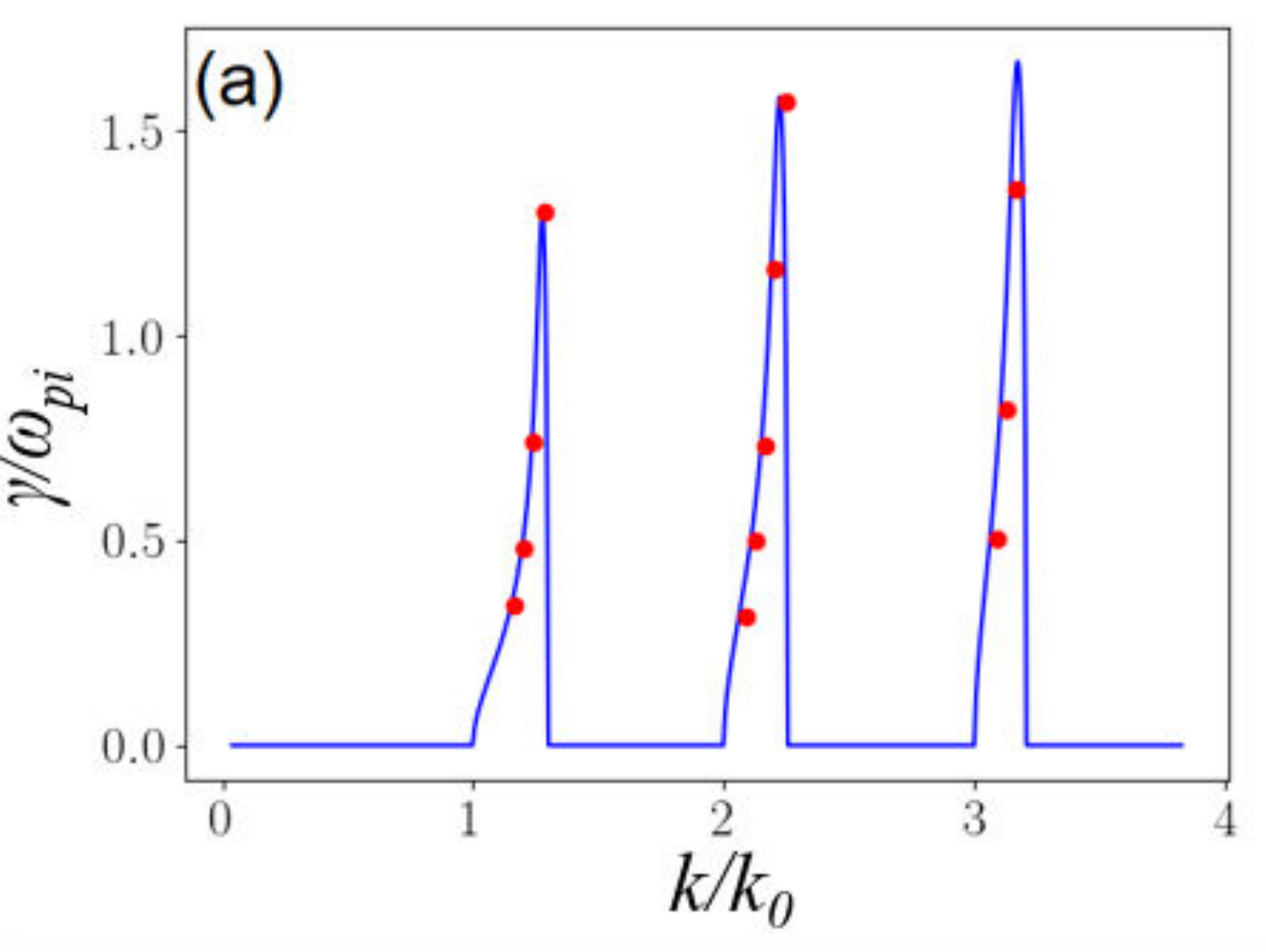}}
\subcaptionbox{\label{fig:t_nek}}{\includegraphics[width=1\linewidth]{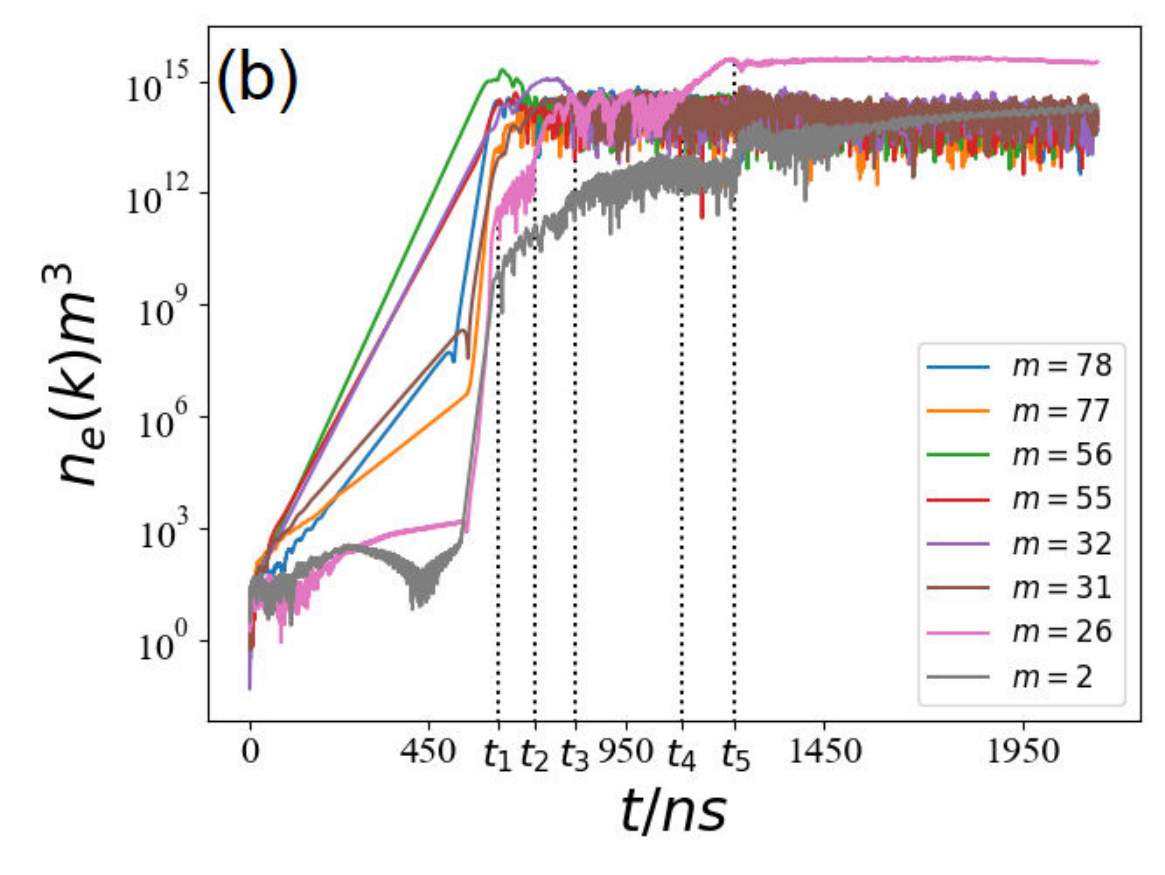}}
\caption{a) Comparison of theoretical growth rates (blue lines) and growth rates found from simulation (red circles).  b) Amplitudes of individual Fourier modes of electron density.}
\end{figure}

In the linear regime, the overlapping resonances of Bernstein and ion-sound modes lead to a resonance condition $\omega/v_d-nk_0- k_x=0$, where $\omega$ is the real frequency, $k_x$ is the wave vector in the $x$-direction, $k_0\equiv \omega_{ce}/v_d$, and $n=1,2,3,\ldots$ shows the number of cyclotron resonances. The full dispersion relation and some of its important limits are discussed in Ref.~\onlinecite{janhunen2018evolution}. To find the linear growth rates, the dispersion relation can be solved by an iterative method as discussed in Ref.~\onlinecite{cavalier2013hall}. For exclusive perpendicular propagation, that is $k_y=0$, the unstable growth rates form discrete bands around the resonance wave vectors $k_x=nk_0$ (see \Cref{fig:growth_rates}). The maximum growth rate of each band does not exactly belong to the corresponding resonance wave vector but to a slightly larger one. On the other hand, from simulation we obtain the amplitude of the individual Fourier modes using the fast Fourier transform (FFT) (\Cref{fig:t_nek}). We note that only the modes $k_x=2m\pi/L$ are resolved by our simulation ($m=1,2,3,\ldots$). In \Cref{fig:t_nek}, two modes around the first cyclotron resonance ($m=31,32$), two modes around the second resonance ($m=55,56$), and two modes around the third resonance ($m=77,78$) are plotted. These modes start to grow linearly after a few nanoseconds, and the linear growth for most of them ends after about 450 ns to 600 ns. The derivative of the amplitude in the linear growth regime represents the growth rate of each mode. These growth rates are compared with the theoretical growth rates in \Cref{fig:growth_rates}.  We note that, due to the steep variations of the unstable growth rates in the 1D configuration, direct growth rate measurements can be highly affected by aliasing or numerical error in the simulations. Therefore, providing data to accurately make this measurement can be challenging for any numerical solver. Nevertheless, \Cref{fig:growth_rates} shows that our low-noise Vlasov solver is capable of reproducing the theoretical growth rates with a reasonable accuracy. 

\begin{figure}[htbp]
\centering
\includegraphics[width=0.5\textwidth]{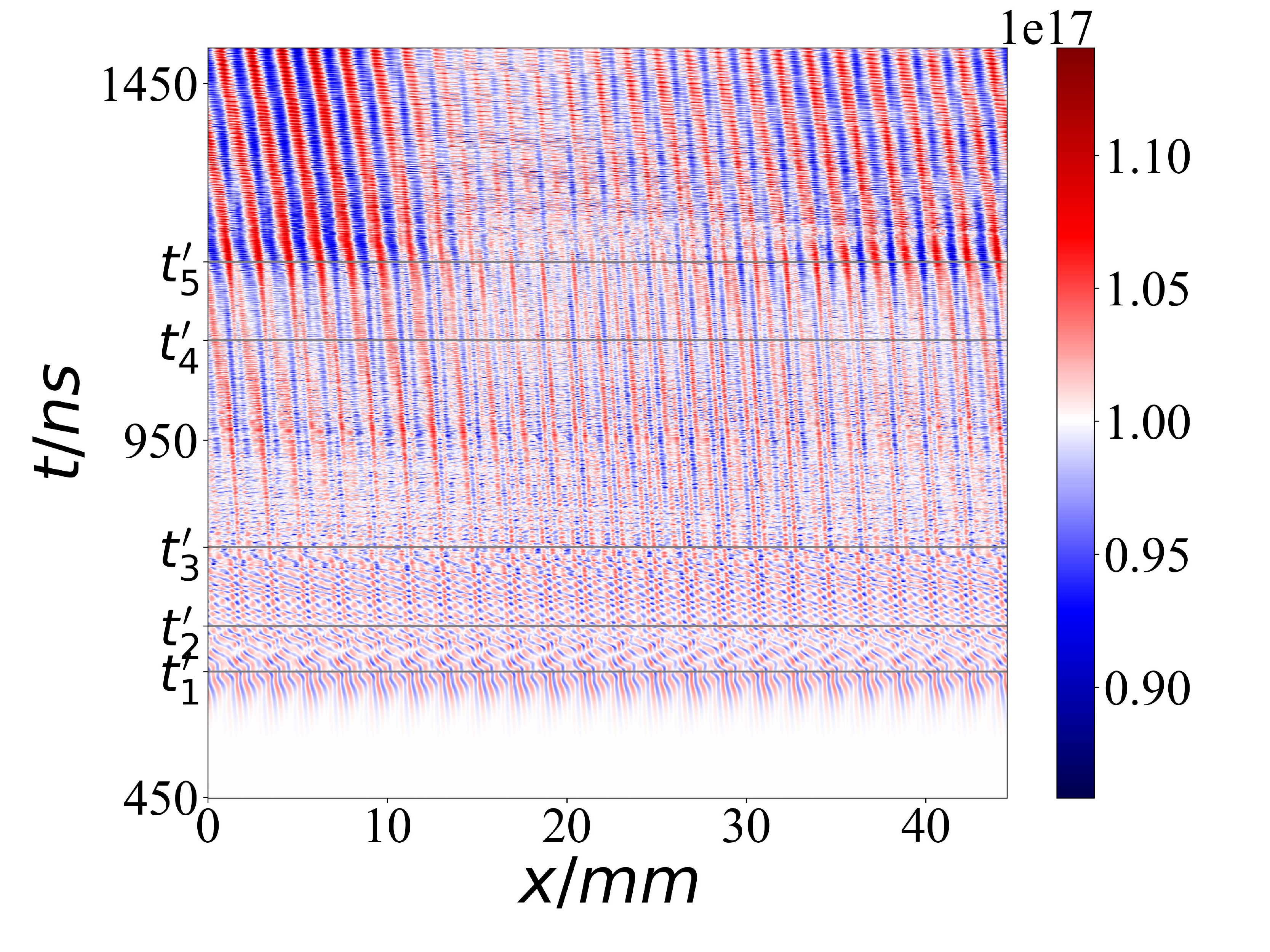}
\caption{The profile of the electron density in  $\text{m}^{-3}$.} 
\label{fig:electron_density}
\end{figure}

\Cref{fig:t_nek} and \Cref{fig:electron_density} show different modes and transitions observed in our simulations. The mode $m=26$ is the dominant mode of the nonlinear regime.  Because its wave vector is close to $k_0$ ($k_0L/2\pi=24.95$), in what follows, we refer to this mode as the ``$k\approx k_0$" mode. The amplitude of this mode remains relatively small in the linear regime and exhibits several transitions between different growth regimes in the nonlinear regime (see \Cref{fig:t_nek}). In \Cref{fig:t_nek}, we have marked five of these transition times with $t_1=626$ ns, $t_2=718$~ns, $t_3=820$ ns, $t_4=1090$ ns, and $t_5=1225$~ns. Also, the mode $m=2$ shows some transitions in  \Cref{fig:t_nek}. This mode does not show any clear growth in the linear regime, whereas its amplitude grows significantly in the nonlinear regime. Several transitions can also be seen in the profile of the electron density in \Cref{fig:electron_density}, where we have marked five of these transition times as $t^\prime_1=626$ ns, $t_2^\prime=690$ ns, $t_3^\prime=800$ ns, $t_4^\prime=1090$ ns, and $t_5^\prime=1200$ ns. We note that these transition times are close to $t_1$ to $t_5$ in \Cref{fig:t_nek}, suggesting a possible correlation between the transitions in electron density and the dominant mode of nonlinear regime. 

In \Cref{fig:electron_density}, for times before $t_1^\prime$, the main observed mode is the dominant mode of the linear regime. Between $t_1^\prime$ to $t_5^\prime$, several transitions can be observed in the wave number and the phase velocity of the dominant modes. A general tendency for transition from shorter wavelengths to the longer wavelengths is observed (the inverse cascade). These transitions can be seen for example at  $t_3^\prime$ and $t_4^\prime$, where some of the preexisting equi-density lines are truncated. At about $t_5^\prime$, we see a transition to the coherent regime (where $k\approx k_0$ is the dominant mode) along with the appearance of some long-wavelength modes with a characteristic size of about the system length ($m=1$ and $m=2$).


\begin{figure}[htbp]
\centering
\captionsetup[subfigure]{labelformat=empty}
\subcaptionbox{\label{fig:Ek}}{\includegraphics[width=1\linewidth]{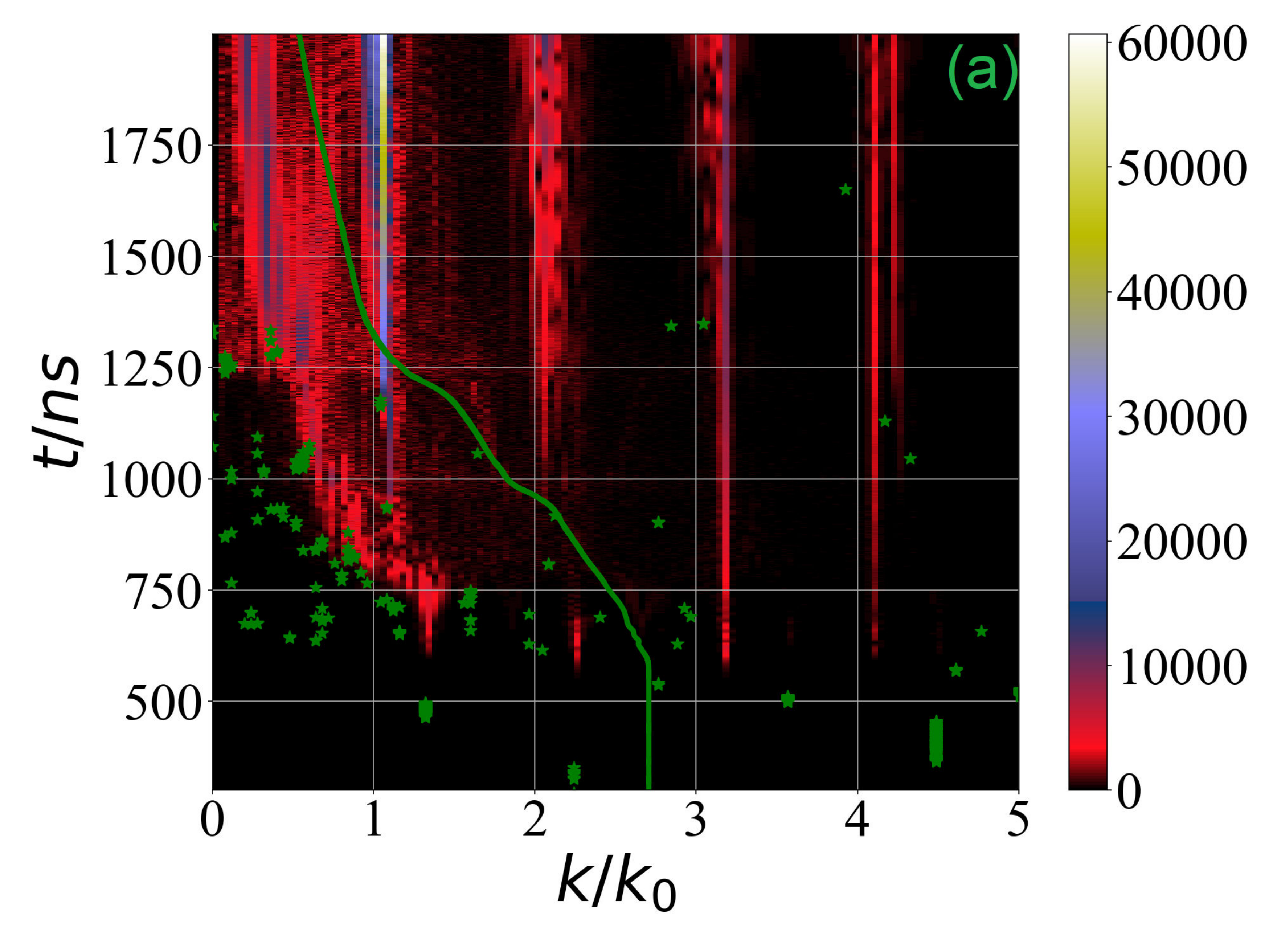}}
\subcaptionbox{\label{fig:gamma_k}}{\includegraphics[width=1\linewidth]{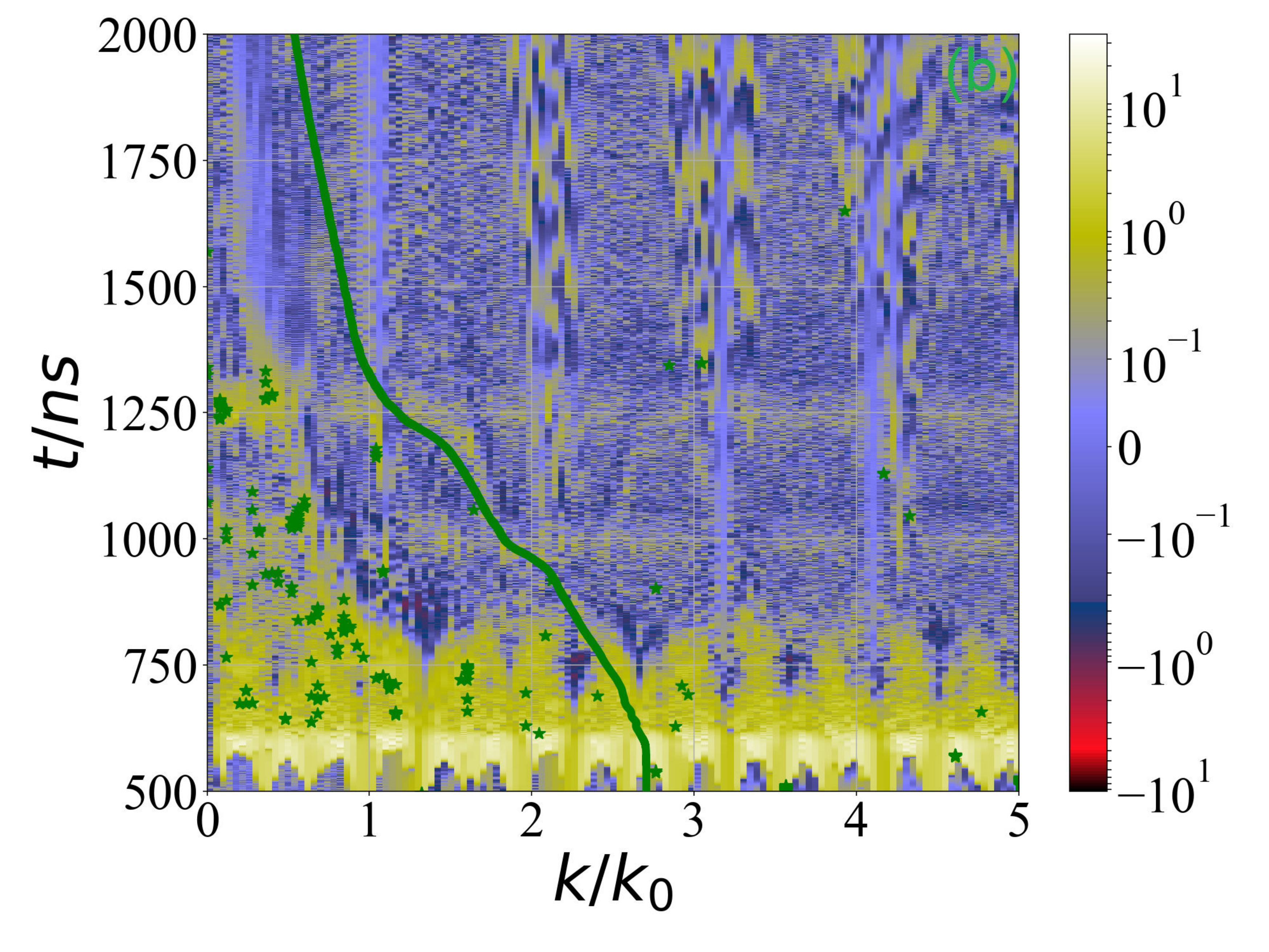}}
\caption{a) $E_k$ in V/m. b) $\gamma_{sim}$ in $\omega_{pi}$. The evolution of the $k^*_{is}$ value is shown by a green line. Green $*$ markers show the $(k,t)$ positions of the most unstable modes in simulations ($k_{sim}^*$); the green color is not related to the mode amplitudes.  \label{fig:Ekgammak}}
\end{figure}

  In \Cref{fig:Ek}, the amplitudes of various Fourier modes of the electric field ($E_k$) are shown. The simulated growth rates, defined by $\gamma_{sim}(k,t)\equiv \dv{\ln E_k}{t}$, are shown in \Cref{fig:gamma_k}. To exclude the fast fluctuations of amplitude, the moving average of $E_k$ (with a window of $3/\omega_{pi}$ in length) is used to calculate $\gamma_{sim}$. In \Cref{fig:t_nek} and at about $t\approx 550$ ns, the modes $m=2,26$ show a fast transition to high amplitudes. \Cref{fig:gamma_k} shows that, at the beginning of nonlinear regime (about $t=600$ ns), most of the modes with a small linear growth rate show a similar transition. An explanation for this transition can be that the low-growth-rate modes are nonlinearly locked to the high-growth-rate modes of the linear regime. For some of the modes, such as $m=2,26$, this transition starts sooner than for others, and the difference in transition start times forms some ``secondary discrete bands" of  high-growth-rate modes, with $k\approx n k_0/3$, at the imminent  nonlinear regime (see $t\approx 500$ ns to $t\approx 550$ ns in \Cref{fig:gamma_k}).

\begin{figure}[htbp]
\centering
\includegraphics[width=0.5\textwidth]{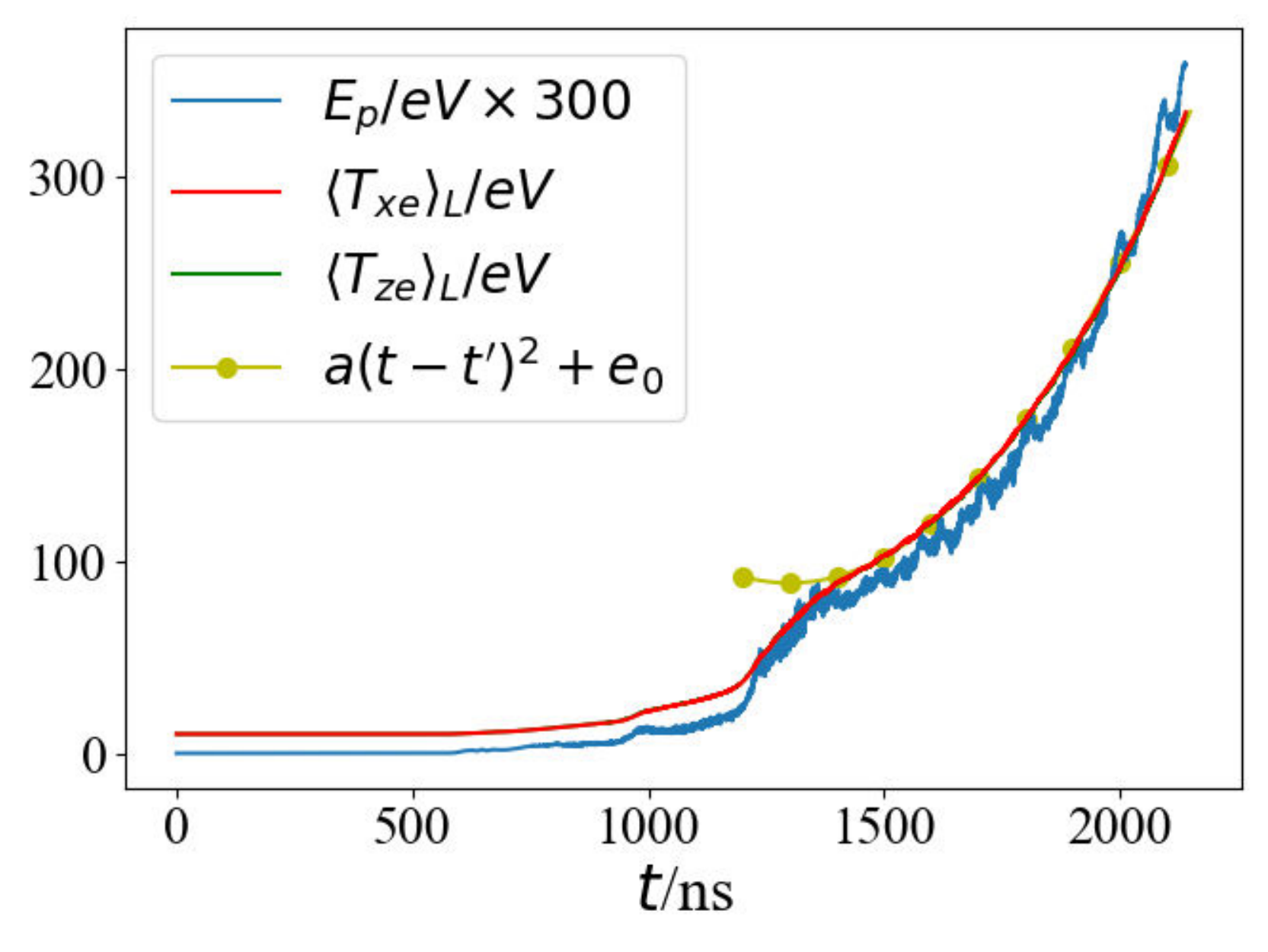}
\caption{The potential energy ($E_p$) and spatially averaged electron temperature along $x$ ($\expval{T_{xe}}$) and $z$ ($\expval{T_{ze}}$). For clarity, $E_p$ is re-scaled by a factor of 300; $t^\prime=1300$ ns, $a=3.4\times 10^{-4}$ eV$/\text{ns}^2$, and $e_0=88.4$ eV.} 
\label{fig:t_ElE}
\end{figure}

In the early nonlinear regime, the inverse cascade can be observed in the spectrum of the electric field and the growth rates in \Cref{fig:Ek,fig:gamma_k}. It is observed that the inverse cascade continues until $t\approx 1200$ ns to $t\approx 1250$ ns. We note that this time approximately coincides with $t_5$, which marks the transition of $k\approx k_0$ to the saturated state (see~\Cref{fig:t_nek}). The quantity $k_{sim}^*(t)$ is defined as the maximum $k$ of the function $\gamma_{sim}$ at each time instant and is also shown in \Cref{fig:Ek,fig:gamma_k}. \Cref{fig:gamma_k} shows that during the inverse cascade, the maximum growth rates mostly belong to low-$k$ modes, leading to an increasing amplitude of these modes in \Cref{fig:Ek}. On the other hand, Ref.~\onlinecite{lampe1972theory} derives the ion-sound growth rate for the nonlinear regime of the ECDI as $\gamma_{is}(k,t)=\qty [\pi/8(m_e/m_i)]^{1/2}kv_d\qty (1+k^2\lambda_D^2)^{-3/2}$. Therefore, the maximum growth rate is expected to occur at $k_{is}^*=\frac{1}{\sqrt{2}\lambda_D}$. Nevertheless, \Cref{fig:Ek,fig:gamma_k} show that $k_{is}^*$ and $k_{sim}^*$ remain far from each other during the nonlinear evolution of ECDI. To calculate $\lambda_D$ in $k_{is}^*$, we have replaced the electron temperature from the simulation (see \Cref{fig:t_ElE}), i.e.,~$\lambda_D(t)=\sqrt{\epsilon_0\expval{T_{xe}}_L/n_0e^2}$, where $\expval{T_{xe}}_L$ is the spatially averaged electron temperature along the $x$-direction.

\begin{figure}[htbp]
\centering
\includegraphics[width=0.5\textwidth]{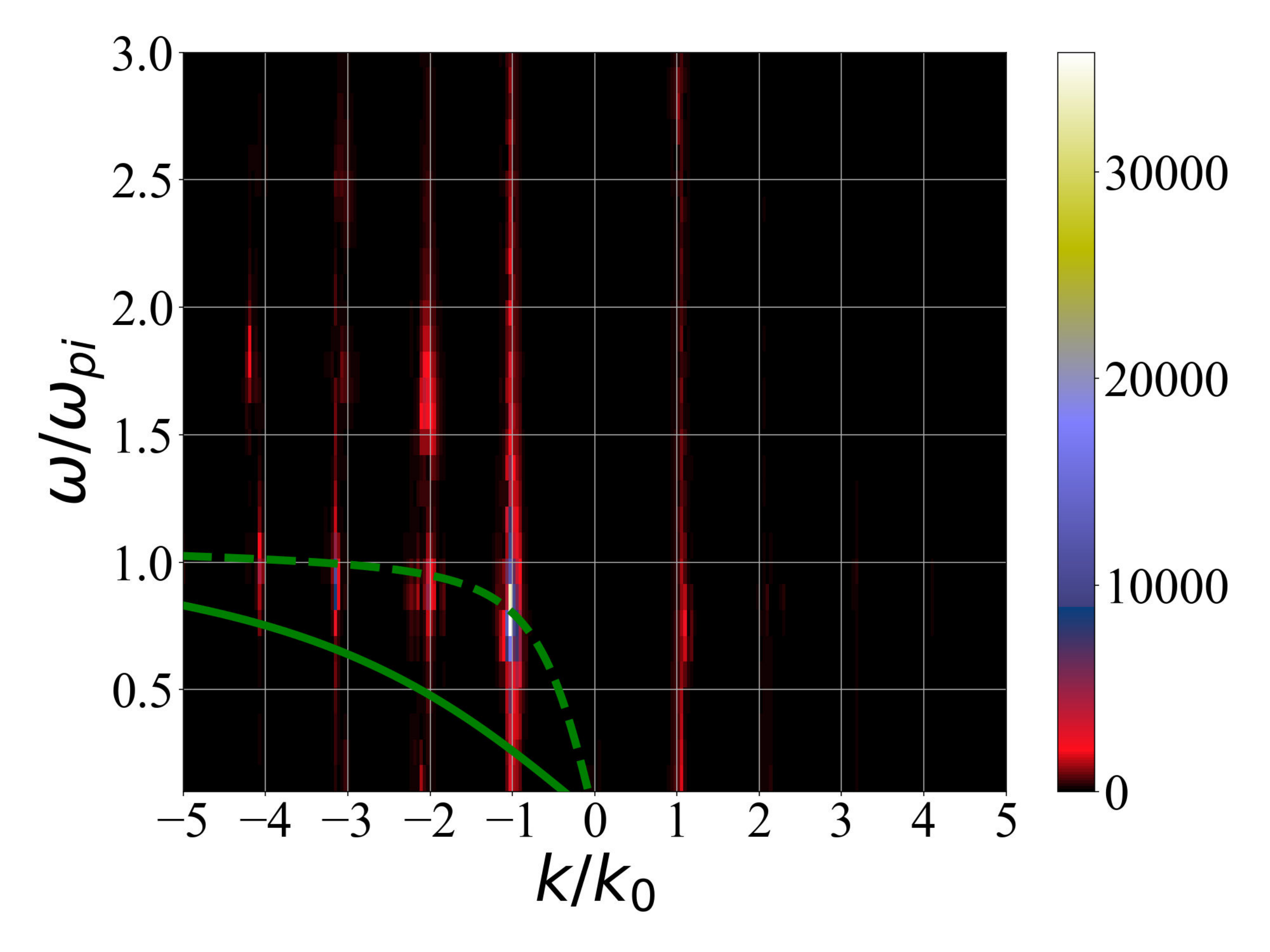}
\caption{The frequency spectrum of $E_x$, in the nonlinear regime. The green solid line and the green dashed line show the ion-sound dispersion relation with the initial temperature and the average temperature at $t=2139$ ns, respectively. In the setup of our problem $k_0\lambda_D=0.2615$.} 
\label{fig:Ekw}
\end{figure}

The frequency spectrum of $E_x$ in the nonlinear regime is shown in \Cref{fig:Ekw}. Similar to the PIC simulations \cite{janhunen2018nonlinear}, the dominant frequencies of $E_x$ are found to be close to $\omega_{pi}$ and its harmonics. Quite different from the ion-sound dispersion relation, the frequency spectrum maintains its discrete feature in the nonlinear regime. The discreteness of the frequency spectrum is also shown in the PIC simulation of Ref.~\onlinecite{janhunen2018nonlinear}. Another point in \Cref{fig:Ekw} is the appearance of the waves moving in the opposite direction of the $E\times B$ drift (backward waves). These waves (also seen at some locations in \Cref{fig:electron_density}) are usually a result of the strong modification of the electron distribution function due to trapping in high-amplitude potential wells of the electric field \cite{tavassoli2021backward,jain2011modeling,HaraPSST2019,DyrudJGR2006}. A similar feature of the backward waves has also been observed experimentally in the Hall thruster plasma\cite{tsikata2009dispersion,tsikata2010three}. 

\begin{figure}[htbp]
\centering
\captionsetup[subfigure]{labelformat=empty}
\subcaptionbox{\label{fig:Jk}}{\includegraphics[width=1\linewidth]{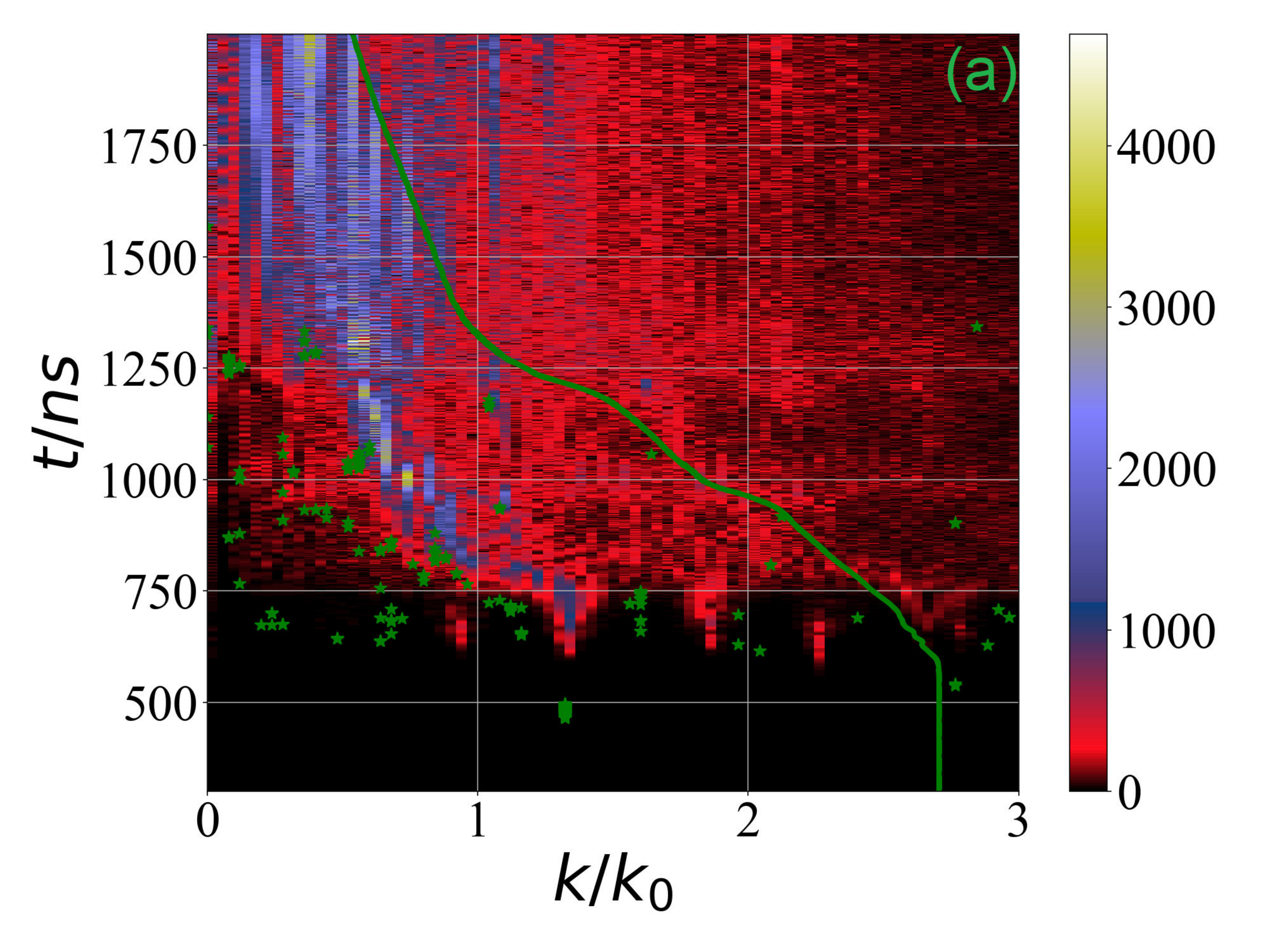}}
\subcaptionbox{\label{fig:t_mvavg_Jze}}{\includegraphics[width=1\linewidth]{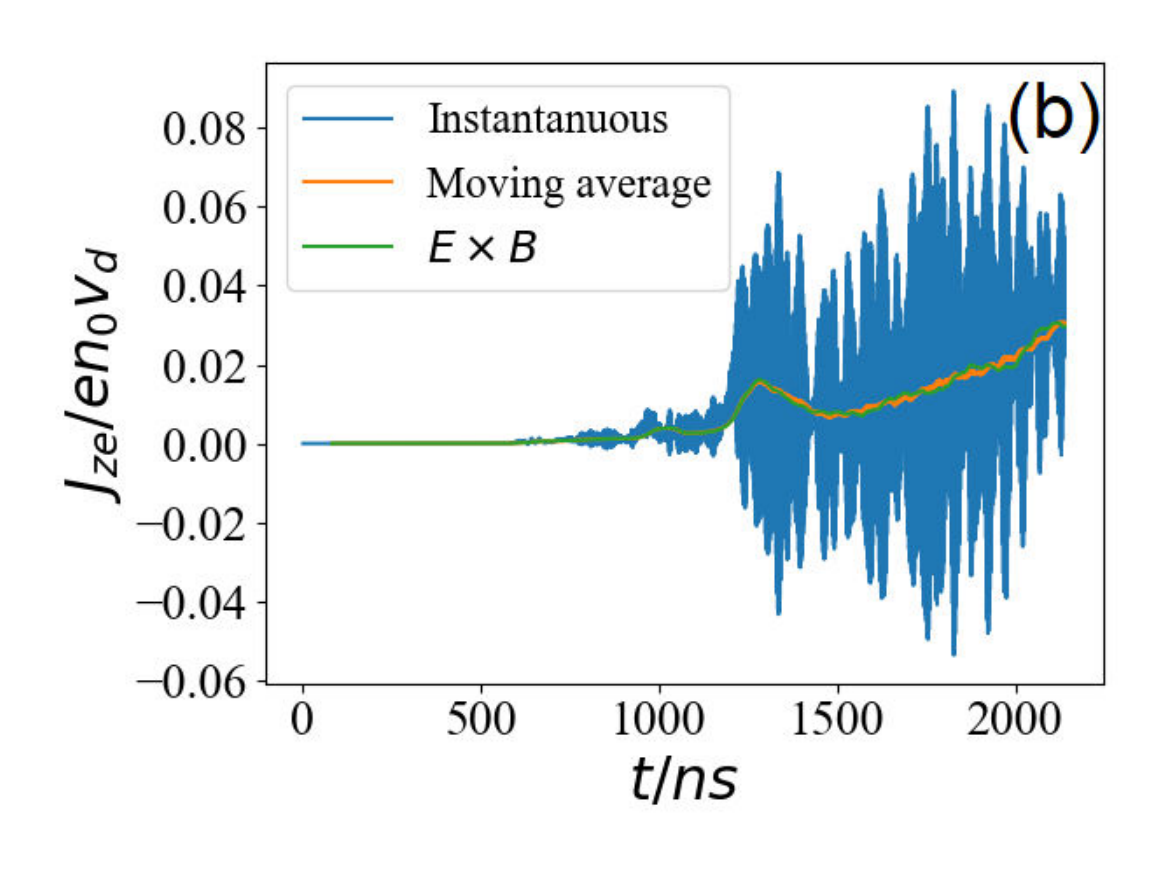}}
\caption{a) $J_k$ in $\text{A}/\text{m}^2$,  b) Comparison of the instantaneous and moving average of the anomalous current with the $E\times B$ anomalous current ($e\expval{n_eE_x}/B_0$). In figure (a), the green line and the green $*$ markers pinpoint the same locations as \Cref{fig:Ekgammak} ($k_{is}^*$ and $k_{sim}^*$).}
\end{figure}

The inverse cascade can also be seen in the spectrum of the anomalous current, $J_k$, in \Cref{fig:Jk}. Due to the inverse energy cascade, intense  modes  are seen in the region $k\lesssim k_0$, similar to previous works \cite{janhunen2018evolution,janhunen2018nonlinear,SmolyakovPPR2020,TaccognaPSST2019}. This is consistent with the notion that  the long-wavelength modes provide the dominant contribution to  anomalous transport. In the low-$k$ region of this figure, $k_{sim}^*$  appears in advance of the intense modes of the anomalous current. This suggests that the intense modes of anomalous current likely resulted from the growth of low-$k$ modes of the electric field. This suggests the inverse cascade in the electric field fluctuations has an important role in the formation of anomalous current. We note that the spectrum of the anomalous current does not show the same coherency as the electric field in \Cref{fig:Ek}, and $k\approx k_0$ ($m=26$) is just one mode among other intense modes. \Cref{fig:t_mvavg_Jze} shows the spatially averaged anomalous current and its moving average. The moving average of the anomalous current is close to the moving average of the $E\times B$ anomalous current, as suggested in Ref.~\onlinecite{lafleur2016theory2} and also confirmed in Ref.~\onlinecite{jimenez_2021}. This observation is in contrast to the PIC simulation of Ref.~\onlinecite{janhunen2018nonlinear}, where the $E\times B$ current is shown to be much smaller than the simulated current. Explaining this discrepancy requires an accurate comparison between the two simulations that takes into account any notable difference in the physical parameters,  numerical parameters, initial conditions, and  post processing; such a comparison is beyond the scope of this study.

\Cref{fig:t_ElE} shows the evolution of the potential energy $E_p\equiv\frac{1}{2n_0L}\int_0^L \epsilon_0E_x^2\,\dd x$ and the spatially averaged electron temperatures. The electron temperature in the $x$- and $z$-directions ($\expval{T_{xe}}_L$ and $\expval{T_{ze}}_L$) are essentially identical during the simulation. In the deep nonlinear regime (after about $t=1300$ ns), the temperatures and potential energy all grow as $t^2$.  The similarity between the growth rates of the temperature and the potential energy is in contrast to those obtained from the ion-sound turbulence theory developed in Refs.~\onlinecite{lampe1971nonlinear,lampe1972anomalous}, where these quantities are expected to have different growth rates. Similar to previous PIC simulations\cite{janhunen2018nonlinear,lafleur2016theory1}, no saturation is obtained in the nonlinear regime. In Ref.~\onlinecite{lafleur2016theory1}, it is suggested that the saturation occurs because of the ion trapping and only when particles are artificially replaced in the simulations when they are displaced  beyond some fixed axial distance  (the virtual axial length model). No such  particle replacement process is present in our Vlasov simulation, although it may exist naturally in azimuthal-axial simulations \cite{BoeufPoP2018,charoy20192d,jimenez_2021,villafana20212d} where particles move out of the region where there is a strong electric field. 

In summary, we investigated the nonlinear regime of the ECDI using a high-resolution Vlasov simulation. This simulation is believed to be free of the statistical noise inherent in  PIC simulations.  A mode with $k\approx k_0$ goes through some transitions before it becomes dominant in the nonlinear regime. Somewhat similar behavior is observed in  the electron density profile.  It is shown that, in the early nonlinear regime, many modes with low linear growth rates show a fast transition to large amplitudes. This growth occurs as a result of a nonlinear locking of these modes to the modes with large linear growth rates, resulting in  fast-growing  secondary (nonlinear) modes.  These transitions are followed by an inverse cascade towards the low-$k$ modes in the nonlinear regime. The inverse cascade terminates at a particular time that is approximately the saturation time of the mode with $k\approx k_0$. The $k_0$ mode remains the dominant electric field mode after this time. As a result of the inverse cascade, the value of the maximum-growth-rate wave vector is much smaller than what is predicted by ion-sound turbulence theory ($k=1/\sqrt(2)\lambda_D$), suggesting that the conditions for the applicability of the ion-sound weak turbulence theory are not likely to be valid in the nonlinear regimes of our simulation. 
One can expect that the resonance broadening due to the nonlinear diffusion will be even weaker in 2D and 3D simulations.  In simulations that involve the direction along the magnetic field, the Modified Two-Stream Instability (MTSI) becomes important \cite{janhunen2018evolution,villafana20212d,TaccognaPSST2019}. It is somewhat surprising that in the latter cases the  linear effects of the finite wave vector along the magnetic field do not annihilate the resonant nature of the ECDI \cite{janhunen2018evolution,villafana20212d,TaccognaPSST2019}, perhaps due to the fact that the most unstable MTSI modes have  long wavelengths. Such effects also depend on the specific parameters of the system, e.g.,~length of the region along the magnetic field.         
Finally, we note that some  observations in our Vlasov simulation are difficult to compare with existing published PIC simulations. Direct head-to-head benchmarking of PIC and Vlasov simulations is suggested for future work.

\section*{Acknowledgment}
This work is partially supported in part by US Air Force Office of Scientific Research FA9550-15-1-0226, the Natural Sciences and  Engineering Council of Canada (NSERC), and Compute Canada computational resources.  

 \section*{Author Declarations}
 \subsection*{Conflict of interest}
 The authors have no conflicts to disclose.
 
 \section*{Data Availability Statement}
 The data that support the findings of this study are available from the corresponding author
upon reasonable request.

\bibliographystyle{apsrev4-2}
\bibliography{Refs.bib}
\end{document}